\documentstyle[graphicx,aps,pre,epsfig]{revtex}

\def\e{{\rm e}}

\begin{document}
\title{Replica bounds for optimization problems and diluted spin systems}

\author{Silvio Franz~\cite{sf}, 
Michele Leone~\cite{ml}}

\address{
\cite{ml} INFM and SISSA, Via Beirut 9, Trieste, Italy \\
\cite{ml,sf} The Abdus Salam International Center for
Theoretical Physics, Condensed Matter Group\\
Strada Costiera 11, P.O. Box 586, I-34100 Trieste, Italy\\
}

\date{\today}
\maketitle

\begin{abstract}
In this paper we generalize to the case of diluted spin models and
random combinatorial optimization problems a technique recently
introduced by Guerra (cond-mat/0205123) to prove that the replica
method generates variational bounds for disordered systems. We analyze
a family of models that includes the Viana-Bray model, the diluted
$p$-spin model or random XOR-SAT problem, and the random K-SAT
problem, showing that the replica method provides an improvable
scheme to obtain lower bounds of the free-energy at all temperatures and
of the ground state energy. In the case of K-SAT the replica method
thus gives upper bounds of the satisfiability threshold.
\end{abstract}


\section{Introduction}
The replica method \cite{EA,PMV}, originally devised as a trick to
compute thermodynamical quantities of physical systems in presence of
quenched disorder, has found applications in the analysis of systems
of very different nature, as Neural Networks, Combinatorial
optimization problems \cite{PMV,comb-opt,nishi}, Error Correction
Codes \cite{nishi} etc.

Although many physicists believe that the method, within the Replica
Symmetry Breaking scheme of Parisi \cite{PMV}, is able to potentially
give the exact solution of any problem treatable as a mean field
theory, the necessary mathematical foundation of the theory is still
lacking, after more then 20 years from its introduction in theoretical
physics. The last times have seen a growing interest of the
mathematical community in the method, leading to important but still
partial results, confirming in certain cases the replica analysis, with
more conventional and well established techniques \cite{matematici}.
Apart the remarkable exception of the analysis of the 
fully connected $p$-spin model in ref. \cite{tala-p-spin} and the
rigorous analysis of Random Energy Models \cite{boko}, the 
analysis of the mathematicians has been, as far as we know, restricted
to the high temperature regions and/or to problem of replica symmetric 
nature. 

Very welcomed have been the techniques recently introduced by Guerra
and Toninelli \cite{guerraTD} which allow rigorous analysis not
relying on the assumption of high temperature, and valid even in
problems with replica symmetry breaking. Along these lines, an important
step towards the rigorous comprehension of the replica method, has
been undertaken in \cite{guerra}, where it has been shown how in
the case of the Sherrington-Kirkpatrick model, and its $p$-spin
generalizations, the replica free-energies with arbitrary number of
replica symmetry breaking steps constitute variational lower bounds to
the true free-energy of the model. As stated in that paper, the
analysis is restricted to fully-connected models, whose replica mean
field theory can be formulated in terms of a single $n\times n$
matrix. However, in recent times, many of the more interesting
problems analyzed with replica theory pertain to the so called
``diluted models'' where each degree of freedom interacts with a
finite number of neighbors. The introduction of a ``population
dynamics algorithm'' \cite{MP-pop} has allowed to treat in full
generality -within statistical precision- complicated sets of
probabilistic functional equations appearing in the one step symmetry
broken framework of diluted models. The same algorithm has been used
as a starting point of a generalized ``belief propagation'' algorithm
for optimization problems \cite{YedidiaLast,MPZ}.  Furthermore,
at the analytic level, simplifications due to graph homogeneities in
some cases \cite{fatt}, and to the vanishing temperature limit in some
other cases \cite{ksat} have led to supposedly exact solutions of the
ground state properties of diluted models, culminated in the
resolution of the random XOR-SAT on uniform graphs in \cite{fatt} and
the random K-SAT problem in \cite{MPZ} within the framework of
``one-step replica symmetry breaking'' (1RSB).

The aim of this paper, is to show that the replica analysis of diluted
models provides lower bounds for the exact free-energy density, and
ground state energy density. We analyze in detail the cases of the
diluted $p$-spin model on the Poissonian degree hyper-graphs also
known as random XOR-SAT problem and the random K-SAT problems. We
expect that along similar lines free-energy lower bounds can be found
for many other diluted cases.

The Guerra method we use sheds some light on the meaning of the
replica mean field theory. The physical idea behind the method is that
within mean field theory one can modify the original Hamiltonian
weakening the strength of the interaction couplings or removing them
partially or totally, and compensate this removal by some auxiliary
external fields. In disordered systems these fields should be random
fields, taken from appropriate probability distributions and possibly
correlated with the original values of the quenched variables
eliminated from the systems. One is then led to consider Hamiltonians
interpolating between the original model and a pure paramagnet in a
random field, and by means of these models achieving free-energy lower
bounds. We will see that the RS case corresponds to assuming
independence between the random fields and the quenched disorder. The
Parisi RSB scheme, assumes at each breaking level a peculiar kind of
correlations, and gives free-energy bounds improving the RS one.

Our paper is organized in this way: in section 2 we introduce some
notations that will be extensively used in the following sections.  In
section 3 we introduce the general strategy to get the replica bounds;
we then specialize to the replica symmetric and the one step replica
symmetry broken bounds, giving the results in the $p$-spin and the
$K$-SAT cases. Conclusions are drawn in section 4. In the appendices some
details of the calculations in both the $p$-spin and the $K$-SAT cases
are shown.

Our results will be issue of explicit calculations. Although at the
end we will get bounds, formalizable as mathematical theorems, the
style and most of the notations of the paper will be the ones
of theoretical physics.

\section{Notations}

The spin models we will consider in this work are defined by a
collection of $N$ Ising $\pm 1$ spins ${\bf S}=\{S_1,...,S_N\}$,
interacting through Hamiltonians of the kind
\begin{equation}
{\cal H}^{(\alpha)}({\bf S},{\bf J})=
\sum_{\mu=1}^M H_{J^{(\mu)}}(S_{i_1^\mu},...,S_{i_p^\mu})
\label{general}
\end{equation}
where the indices $i_l^\mu$ are i.i.d.  quenched random variables
chosen uniformly in $\{1,...,N\}$.  We will call each term
$H_{J^{(\mu)}}$ a clause.  The subscript $J^{(\mu)}$ in the clauses
indicates the dependence on a single or a set of quenched random
variables, as it will be soon clear.  The number of clauses $M$ will
be taken to be proportional to $N$. For convenience we will choose it
to be for each sample a Poissonian number with distribution
$\pi(M,\alpha N)=\e^{-\alpha N}\frac{(\alpha N)^M}{M!}$. The
fluctuations of $M$ will not affect the free-energy in the
thermodynamic limit, and this choice, which  slightly simplify the
analysis, will be equivalent to choosing a fixed value of $M$ equal to
$\alpha N$.
The clauses themselves will be random.  The $p$-spin
model\cite{ri-we-ze} has clauses of the form
\begin{equation}
H_{J^{(\mu)}}(S_{i_1^\mu},...,S_{i_p^\mu})=J^\mu
S_{i_1^\mu}\cdot ...\cdot S_{i_p^\mu} \; .
\label{spin}
\end{equation}
This form reduces to $H_{J^{(\mu)}}(S_{i_1^\mu},S_{i_2^\mu})=J^\mu
S_{i_1^\mu}S_{i_2^\mu}$ in the case of the Viana-Bray spin glass
$p=2$.  In both cases the $J^\mu$ will be taken as i.i.d. random
variable with regular symmetric distribution $\mu(J)=\mu(-J)$. Notice
that for $\mu(J) = 1/2 [ \delta (J+1) + \delta (J-1)]$ the model
reduces to the random XOR-SAT problem \cite{XOR-SAT} of computer
science.  The random K-SAT clauses have the form \cite{ksat}
\begin{equation} 
H_{J^{(\mu)}}(S_{i_1^\mu},...,S_{i_p^\mu})=
\prod_{l=1}^p \frac{1+J^\mu_{i_l^\mu}S_{i_l^\mu}}{2} \; ,
\label{SAT}
\end{equation} where the
$J^\mu_{i_l^\mu}=\pm 1$ are i.i.d. with symmetric probability.  (The
number $p$ of spin appearing in a clause is usually called $K$ in the
K-SAT problem, for uniformity of notation we will deviate from this
convention).  Notice that in all cases, on average each spin
participate to $\alpha = \frac M N$ clauses, and that the set of spins
and interactions defines a random diluted hyper-graph
of uniform rank $p$ and random local degree with Poissonian statistics
in the thermodynamic limit. At high enough temperature, the existence of 
the free-energy in the
thermodynamic limit for models of this kind has been proved in by
Talagrand in \cite{tala-ksat}, together with the validity of the RS
solution. A proof valid at all temperature based on the ideas
presented in this paper, can be obtained for even $p$
in analogy of the analysis in \cite{guerraTD} for long range
models. We sketch it in appendix C in the case of the $p$-spin model.

In establishing the free-energy bounds we will need several kind
of averages:
\begin{itemize}
\item The Boltzmann-Gibbs average for fixed 
quenched disorder: given an observable $A({\bf S})$
\begin{equation}
\omega(A)=\frac{\sum_{{\bf S}}A({\bf S}) \exp(-\beta {\cal H}({\bf S},{\bf
J}))}{Z}
\end{equation}
where $Z={\sum_{{\bf S}}\exp(-\beta {\cal H}({\bf S},{\bf J}))}$ and 
$\beta$ is the inverse temperature. 

Obviously, $\omega(A)$, as well as $Z$ will be functions of the quenched
variables, the size of the system and the temperature. This dependence will
be  made explicit only when needed. 
\item
The disorder average:
given an observable quantity $B$
dependent on the quenched variables appearing in the Hamiltonian, 
we will denote as $E(B)$ its average. This will include the average
with respect to the $J$ variables and the choice 
of the random indices in the clauses as well as with respect to other 
quenched variables to be introduced later. 
\item
We will need in several occasion the ``replica measure'' 
\begin{equation}
\Omega (A_1,...,A_n)=E( \omega(A_1)...\omega(A_n))
\label{repmeas}
\end{equation}
and some generalizations that we will specify later. 
\item
We will occasionally use other kinds of averages, as well as other
notations, for which we will use an angular bracket notation, with 
a subscript indicating the variable(s) over which the average is
performed. e.g. an average over a random variable $u$ with probability
distribution $Q (u)$ will be denoted equivalently as
$\int d u Q (u) ( \cdot )  \equiv  \int dQ (u) ( \cdot ) 
\equiv \langle \cdot \rangle_u$. 
Analogously, averages over distribution
families of $Q(u)$ will be denoted as
$\int d Q  {\cal Q} (Q) ( \cdot ) \equiv \int {\cal D Q} (Q) ( \cdot ) 
\equiv \langle \cdot \rangle_{Q}$. Subscripts will be
omitted whenever confusion is not possible.
\item
Another notation we will have the occasion to use in the one 
for the overlaps among $l$ spin configurations $\{ S_i^{a_1},...,S_i^{a_l} \}$,
out of a population of $n$  
$\{ S_i^{1},...,S_i^{n} \}$: 
\begin{equation}
q^{(a_1,...,a_l)}=\frac 1 N \sum_{i=1}^N S_i^{a_1} \cdot ... \cdot
S_i^{a_l} \;\;\;\; (1\le a_r \le n \;\;\; \forall r) ,
\end{equation} 
and in particular 
\begin{equation}
q^{(n)}=q^{(1,...,n)}=\frac 1 N \sum_{i=1}^N S_i^{1} \cdot ... \cdot S_i^{n} \; ,
\end{equation} 
This notation will be extended to multi-overlaps in the 1RSB case, as we will
specify in section \ref{tarta}.
\end{itemize}
In the following we will need to consider averages where some of the 
variables are excluded, e.g. the averages when a variable 
$u_i^{k_i}$ is erased. 
These average will be denoted with a subscript
$-u_i^{k_i}$ e.g. if an $\omega$ average is concerned the notation 
will be $\omega(\cdot)_{-u_i^{k_i}}$. 
Other notations will be defined later in the text whenever needed. 

Our interest will be confined to bounds to the free-energy density 
$F_N = -\frac{1}{\beta N} E \log Z$ and the ground state energy density $U_{GS} = 
\lim_{N \to \infty} 1/N E \left[ min \, (U_N) \right] $ valid in the thermodynamic
limit, so that $O(1/N)$ will be often implicitly neglected in our 
calculations.

\section{The general strategy} 

The strategy to get the replica bound is a generalization of the one
introduced by Guerra in the case of fully connected models
\cite{guerra}. We will consider models which will interpolate between
the original ones we want to analyze and pure paramagnet in random
fields with suitably chosen distribution.  The underlying idea is that,
given the mean field nature of the models involved,
if one was able to reconstruct the real local fields acting on a given
spin variable via a given hyper-edge, and to introduce auxiliary fields
acting on that variable in such a way to energetically balance
the deletion of the hyper-edge, then it would be possible to have an exact
expression for the free-energy in terms of such auxiliary fields even
when the whole edge set was emptied. However, if the replacement is
done with some approximate form of the auxiliary fields distribution
function, the real free-energy will be the one calculated using the
approximate fields plus an excess term at every step of the graph
deletion process. The proof of the definite sign of this excess term
gives a way to determine bounds for the thermodynamic
quantities. 

We will prove the existence of replica lower bounds to the free-energy
density of the p-spin model and the random K-SAT problem. In this last
case our result proves that the recent replica solution of \cite{MPZ}
gives a lower bound to the ground state energy and therefore an
upper bound for the satisfiability threshold.  The proofs will strictly
hold in the $N \to \infty$ limit, due to the presence of corrections
of order $1/N$ in the calculated expressions for any finite size
graph.  Moreover, our proofs will be restricted to 
the 
$p$-spin model the the K-SAT with even $p$. In the cases of odd $p$ 
the same bound would hold if one could rely on some physically reasonable 
assumptions on the overlap distribution (see below).

\begin{figure}
\centering
\epsfxsize=0.4\textwidth \epsffile{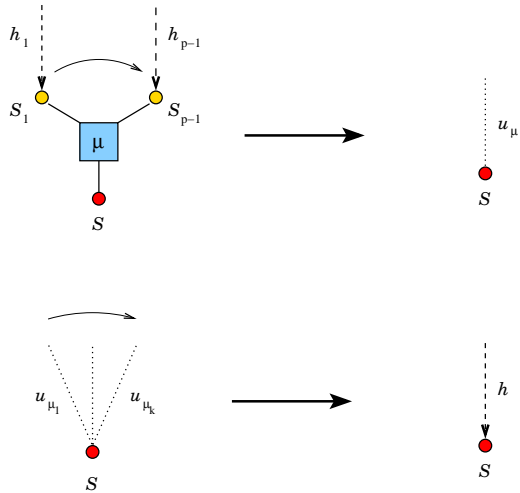}
\caption{Diagrammatic representation of the cavity relations for $h$ and $u$ fields acting
on spin $S$. The hyper-edge
interaction is drawn in the factor-graph notation.}
\label{cavity}
\end{figure}

Our analysis will start from the TAP equations for the models
\cite{naka,MP-pop}, and their probabilistic solutions implied by
the cavity, or equivalently the replica method at various degrees of
approximation. We will consider in particular the replica symmetric
(RS) and one step replica symmetry broken solutions, but it should be
clear from our analysis how to generalize to more steps of replica
symmetry breaking. In the TAP/cavity equations one singles out the
contribution of the clauses and the sites to the free-energy and
defines cavity fields $h_i^{(\mu)}$ and $u_\mu^{(i)}$ respectively as
the local field acting on the spin $i$ in absence of the clause $\mu$
and the local field acting on $i$ due to the presence of the clause
$\mu$ only. If we define $Z_N[S_i]$ as the partition function of a
given sample with $N$ spins where all but the spin $i$ are integrated,
$F_{N,-i}$ the free-energy of the corresponding systems where the spin
$S_i$ and all the clauses to which it belongs are removed, we can
write,
\begin{eqnarray}
Z_N[S_i]& = &\e^{-\beta F_{N,-i}}\prod_{\mu\in T_i}
\sum_{S_{i_2^\mu},...,S_{i_p^\mu}} \e^{-\beta
H_{J^{(\mu)}}(S_{i^\mu},S_{i_2^\mu},...,S_{i_p^\mu})+\sum_{l=2}^{p}
h_{i_l^\mu}^{(\mu)}S_{i_l^\mu} }
\nonumber\\ 
& = & \e^{-\beta
F_{N,-i}}
\prod_{\mu\in T_i} 
B_{\mu}^{(i)} 
\e^{\beta u_\mu^{(i)}S_i}
\label{c1}
\end{eqnarray}
where $T_i$ is the set of clauses containing the spin $i$, and
the constant $B_{\mu}^{(i)}=\e^{-\beta
\Delta F_{\mu}^{(i)}}$ 
can be interpreted as suitable shifts in the free-energy due to the
contribution of the clause $\mu$ for fixed value of the spin $i$. 
We notice that denoting $J^{\mu}$ as $J$, and renaming the fields 
in (\ref{c1}) into $h_1,...,h_{p-1}$,
Eq. (\ref{c1}) defines functions 
\begin{equation}
u_J(h_1,...,h_{p-1}) \; \;  {\rm and} \; \;
B_J(h_1,...,h_{p-1}) \; .
\label{ub}
\end{equation} 
The equation are closed by the self-consistent condition: 
\begin{equation}
h_{i}^{(\mu)}=\sum_{\nu \in \{ T_i-\mu\}} u_\nu^{(i)}
\label{c2}
\end{equation}
These equations are at the basis of iterative algorithms such as the
``belief propagation'' or ``sum-product'' know for a long time in
statistical inference \cite{pea} and coding theory \cite{galla}
and the more recently proposed algorithm of ``survey
propagation''\cite{MPZ}.  Conditions (\ref{c1}) and (\ref{c2}) can be
diagrammatically represented as in fig.(\ref{cavity}).
The
cavity fields solutions of (\ref{c1},\ref{c2}) are random variables
which fluctuate for two reasons \cite{PMV,MP-pop,MP}. 
First, they differ from sample to
sample. Second, within the same sample the equations can have several
solutions which can level-cross. The cavity/replica method provides
under certain assumption probabilistic solutions. In the RS
approximation, one just supposes a single solution to give the 
relevant contribution in a given
sample. The sample to sample fluctuation induce probability
distributions $P(h)$ and $Q(u)$ whose relations implied by
(\ref{c1},\ref{c2}) are:
\begin{eqnarray}
P(h)& = &\sum_k \e^{-\alpha p}\frac{(\alpha p)^k}{k!}
\int du_1 \;Q(u_1) ... du_k \;Q(u_k) 
\delta (h - \sum_{i=1}^k u_k)\\
Q(u)& = & \int dh_1 \;P(h_1)... dh_{p-1} \;P(h_{p-1})
\langle \delta (u-u_J(h_1,...,h_{p-1}))\rangle_J
\end{eqnarray}
where $\langle \cdot\rangle_J$ denotes the average over the random
variables appearing in a clause.
In addition to sample to sample fluctuations, the 1RSB solution
assumes fluctuations of the fields from solution to solution of the
equations, so that the functions $P(h)$ and $Q(u)$ will be themselves
randomly distributed according to some functional probability
distributions ${\cal P}(P)$ and ${\cal Q}(Q)$ related by the
self-consistency equations \cite{remi}
\begin{equation}
{\cal Q}(Q)=\int {\cal D}P_1 {\cal P}(P_1) ... {\cal D}P_{p-1} {\cal
P}(P_{p-1}) \langle \delta (Q(\cdot)-
Q(\cdot|P_1,...,P_{p-1},J))\rangle_J
\label{probQ1}
\end{equation}
\begin{equation}
{\cal P}(P) = \sum_{k=0}^\infty \e^{-\alpha p}\frac {(\alpha p)^k}{k!} 
\int \prod_{l=1}^{k} {\cal D}{Q_l}
{\cal Q}({Q_l})
\delta(P(\cdot)- P(\cdot|{Q_1},...,{Q_k} ) )
\label{ProbP1}
\end{equation}
 where:
\begin{eqnarray}
& & Q(u|P_1,...,P_{p-1},H) =  {\cal N}_P [P_1,...,P_{p-1}] \int dh_1\; P_1(h_1) 
...dh_{p-1}\; P_1(h_{p-1}) B_{\bf J}(h_1,...,h_{p-1})^m
\delta (u-u_{\bf J}(h_1,...,h_{p-1}) ) \\
& & P(h|{Q_1},...,{Q_k} ) =  {\cal N}_{Q,k}[{Q_1},...,{Q_k}] \, \, (2\cosh(\beta h))^m \int \prod_{l=1}^k
d u_l \; \frac{ {Q_l} (u_l)}{(2\cosh(\beta u_l))^m} 
\delta ( h - \sum_{l=1}^k u_l )
\label{innerprobs1}
\end{eqnarray}
where ${\cal N}_{Q,k}[{Q_1},...,{Q_k} ]$ and ${\cal N}_G
[G_1,...,G_{p-1}]$ insure normalization and $B_{\bf
J}(g_1,...,g_{p-1})$ is a rescaling term of the form (\ref{ub}) that
can be reabsorbed in the normalization in the case of the $p$-spin
model.  Its form for the $K$-SAT case is given in the appendix.  $m$
is a number in the interval $(0,1]$, which within the formalism
selects families of solutions at different free-energy levels. The
physical free-energy is estimated maximizing over $m$.

The interpretation of these equations has been discussed many times
in the literature \cite{PMV,MP-pop,MP}. We will show here, that such
choices in the
field distributions result in lower bounds for the free-energy
analogous to the ones first proved by Guerra in fully connected models. 
In order to prove these bounds, we will have
to consider auxiliary models where the number of clauses 
$\alpha N$ will be reduced to
$\alpha t N$ ($0\leq t\leq 1$), while this reduction will be compensated
in average by some external field terms of the kind:
\begin{equation}
{\cal H}^{(t)}_{ext} =  \sum_i \sum_{l_i=1}^{k_i} u_i^{l_i} S_i
\end{equation}
where the numbers $k_i$ will be i.i.d. Poissonian variables with
average $\alpha p (1-t)$.  As the notation suggests, the fields
$u_i^l$ will play the role of the cavity fields $u_\mu^{(i)}$ of
the TAP approach, and they will be i.i.d. random variables with suitable
distribution. Indeed, for each field $u_i^{l_i}$ we will chose 
in an independent way 
$p-1$ primary fields $g_i^{l_i,n}$ ($n=1,...,p-1$) and clause variables 
$J_i^{l_i,n}$ such that the relation
\begin{equation}
u_i^{l_i}=u_{J_i^{l_i,n}}(g_i^{l_i,1}...g_i^{l_i,p-1})
\end{equation}
is verified.
Notice that the compound Hamiltonian 
\begin{equation}
{\cal H}_{tot}^{(t)}[{\bf S}]={\cal
H}^{(\alpha t)}[{\bf S}]+{\cal H}^{(t)}_{ext}[{\bf S}]
\label{compoundH}
\end{equation}
 will constitute a sample with
the original distribution for $t=1$, while it will consist in a system
of non interacting spins for $t=0$.  The key step of the procedure,
consists in the choice of the distribution of the primary fields
$g_i^{l_i}$. We will also find useful to define fields $h_i$ verifying 
\begin{equation}
h_i= \sum_{l=1}^{k_i} u_i^l. 
\end{equation}
The field $u$ are related to the $g$'s by a relation similar to
(\ref{c1}), while the $h$'s are related to the $u$'s by a relation
similar to (\ref{c2}). Of course, the statistics of the fields $h$ and
the $g$'s do coincide in the TAP approach. It is interesting to note
that the bounds we will get, are optimized precisely when their
statistical ensemble coincide. As we mentioned, various Replica bounds
are obtained assuming for the fields $g_i^{l_i}$ the type of
statistics implied by the different replica solution. So, the Replica
Symmetric bound is got just supposing the field as quenched variables
completely independent of the quenched disorder and with distribution
$G(g)$.  For the one-step RSB bound on the other hand the distribution
$G$ will itself be considered as random, subject to a functional
probability distribution ${\cal G}[G]$. More complicated RSB
estimates, not considered in this paper, can be obtained along the
same lines.  The case of the fully connected models considered by
Guerra can be formalized in this way where the various field
distributions involved are Gaussian.

\subsection{The RS bound}

We consider in this case i.i.d. fields $u$ and $h$ distributed
according probabilities $Q(u)$ and $P(h)$ verifying the following
relation with the distribution $Q(g)$ of the primary fields.
\begin{eqnarray}
Q(u)& = & \int dg_1 \;G(g_1)... dg_{p-1} \;G(g_{p-1})
\langle \delta (u-u_J(g_1,...,g_{p-1}))\rangle_J\\
P(h)&=&P(h|k )\pi(k,\alpha p (1-t))\\
P(h|k )& = &\int du_1 \;Q(u_1) ... du_k \;Q(u_k) 
\delta (h - \sum_{i=1}^k u_k)
\end{eqnarray} 
The distribution $G(g)$ will be chosen to be symmetric under change of
sign of $g$, and regular enough for all the expression below to make
sense. The RS bound can now be obtained following a procedure to the
one of Guerra for the SK model, and considering the $t$ dependent
free-energy; with obvious notation:
\begin{equation}
F(t)=\lim_{N\to\infty} F_N(t)=\lim_{N\to\infty}-\frac{1}{\beta N} E \log Z_N(t)
\end{equation} 
where $E$ represents the average over all the quenched variables, the
one defining the clauses and the external fields. We then consider
the $t$ derivative of $F_N$
\begin{equation}
 \frac {d}{d t}F_N(t)= -\frac{1}{N\beta } \frac {d}{dt} E (\log Z_N). 
\end{equation}
As in \cite{guerra} we will then write
\begin{equation}
F(1)=F(0)+\int_0^1 d t \; \frac {d}{dt}F(t)
\label{quattordici}
\end{equation}
and show, by an explicit computation, that, up to $O(1/N)$ terms that
will be systematically neglected, the expression coincides with the
variational RS free-energy plus a remainder. In fortunate cases this
term will have negative sign and neglecting it  will immediately
result in a lower bound for the free-energy. This happens in the
Viana-Bray model, the $p$-spin and the K-SAT for even $p$. In the
cases of odd $p$ we were not able to prove the sign definiteness of
the remainder, although as we will discuss we believe this to be the
case on a physical basis.

The time derivative of $F$ take contributions from the derivative of
the distribution of the number of clauses $M$
\begin{equation}
\frac {d \pi(M,\alpha t N)}{d t}=-N\alpha (\pi(M,\alpha t N)-
\pi(M-1,\alpha t N))
\end{equation}
and the distribution of 
the number of $u$ fields on each site 
\begin{equation}
\frac {d \pi(k_i,\alpha p(1-t))}{d t}=\alpha p(\pi(k_i,\alpha
p(1-t))-
\pi(k_i-1,\alpha p(1-t)))
\end{equation}
so that:
\begin{eqnarray}
\frac {d}{dt} E \log Z(t)= & & -N\alpha \sum_M (\pi(M,\alpha t
N)-\pi(M-1,\alpha t N)) E' \log Z(t)
\nonumber
 \\
& &+\alpha p \sum_i \sum_{k_i}
(\pi(k_i,\alpha p(1-t))-\pi(k_i-1,\alpha p(1-t))) E''_i  \log Z(t)
\label{der}
\end{eqnarray}
where we have denoted as $E'$ the average with respect to all the 
quenched variables except $M$ and with $E''_i$ the average with
respect to all the quenched variables except $k_i$, and simply $Z(t)$
the partition function of the $N$ spin system $Z_N(t)$. 

In the first term of (\ref{der}) we can single out the $M$-th clause,
and write $Z(t)=Z_{-M}(t)\omega(\e^{-\beta
H_M(S_{i_1^M},...,S_{i_p^M}})_{-M}$, where by $Z_{-M}(t)$ we denote the
partition function of the system in absence of the $M$-th clause, and
$\omega(\cdot)_{-M}$ is the canonical average
in absence of the $M$-th clause. 
In the following
terms we single out the $k_i$-th field $u$ term,
$Z(t)=Z_{-u_i^{k_i}}(t) \omega(\e^{\beta u_i^{k_i}S_i})_{-u_i^{k_i}}$,
where $Z_{-u_i^{k_i}}(t)$ is the partition function in absence of the
field ${-u_i^{k_i}}$ and analogously for the average
$\omega(\cdot)_{-u_i^{k_i}}$. 
Finally, rearranging all terms we find
\begin{eqnarray}
\frac {d}{dt} E \log Z(t)= & &N\alpha \sum_M (\pi(M-1,\alpha t N))
E' \log [\omega(\e^{-\beta
H_{J^{(M)}}(S_{i_1^M},...,S_{i_p^M})})_{-M})]
\nonumber
\\& &
- p \alpha \sum_i\sum_{k_i} \pi(k_i-1,\alpha p(1-t))
E''_i\log[\omega(\e^{\beta u_i^{k_i}S_i})_{-u_i^{k_i}}].
\label{pippo}
\end{eqnarray}
where we have used $\sum_M \pi(M-1,\alpha t N) E' \log Z_{-M}=
\sum_{k_i} \pi(k_i-1,\alpha p(1-t))E''_i\log Z_{-u_i^{k_i}}= E\log
Z$.  We notice at this point that the statistical ensemble defined by
$\pi(M-1,\alpha t N)) E'$ can be substituted with the original one
$E$ and the average of the variables appearing in the clause we have 
singled out. To be more precise, we remark that the average $\omega(\cdot)$
depends on the quenched variables $D=\{ {\bf J},{\bf u}\}$ appearing
in the Hamiltonian. Writing explicitly this dependence as
$\omega(\cdot|D)$, and denoting as $D_{-M}$ all the quenched variables  
except the ones appearing in the $M$-th clause, our statement is that 
thanks to the Poissonian distribution of $M$ and the uniform choice of
the indices of each clause, 
\begin{equation}
\sum_M (\pi(M-1,\alpha t N)) E'
\log [\omega(\e^{-\beta H_{J^{(M)}}(S_{i_1^M},...,S_{i_p^M})}|D_{-M})]=
E\left( 
\frac{1}{N^p} \sum_{i_1,...,i_p} \langle
  \log [\omega(\e^{-\beta H_J(S_{i_1},...,S_{i_p})}|D)]\rangle_J
\right).
\end{equation}
where by $\langle \cdot \rangle_J$ we denote the average with respect
to the random variables appearing in the clause. This is a crucial
step in our analysis, in fact, 
similar considerations apply to the term in the second line of (\ref{pippo}),
which can be written as
\begin{eqnarray}
\sum_{k_i} \pi(k_i-1,\alpha p(1-t)) E''_i\log[\omega(\e^{\beta
u_i^{k_i}S_i})_{-u_i^{k_i}}] =
& & E \left\langle \log \omega \left(e^{\beta u S_i}\right) \right\rangle_u \; .
\end{eqnarray} 
The same kind of averages $E$ and $\omega$ appear in the two terms
which can be therefore directly compared as we will do in the next
section. This property, linked to the Poissonian character of the
graph defined by the model would not hold for other ensembles of
random graphs and the analysis would be technically more involved.
Substituting in (\ref{pippo}) we find:
\begin{equation}
\frac{1}{N}\frac {d}{dt} E \log Z(t)=\alpha E \left[
\frac{1}{N^p} \sum_{i_1,...,i_p} \langle
  \log [\omega(\e^{-\beta H(S_{i_1},...,S_{i_p})})]\rangle_H
-\frac{p}{N}\sum_i
\langle \log \omega (e^{\beta u S_i}) \rangle_u \right] 
\label{pippo2}
\end{equation}  
Rearranging terms and using 
(\ref{quattordici}) we finally find that 
the free-energy $F_N$ can be written as 
\begin{equation}
F_N = F_{var}[G] + \int_0^1 dt\; R_{RS}[G,t]+O(1/N)
\label{FveraRS}
\end{equation} 
where $F_{var}[G]$ coincides the expression of the variational free-energy in
the replica treatment under condition
$G[h] = P[h]$ $\forall \; h$ at $t=0$ and $\int_0^1 dt\; R_{RS}[G,t]$ is a remainder term.
Instead of writing the formulae
for general clauses, in order to keep the notations within reasonable
simplicity, we specialize now to the specific cases of the $p$-spin model and
the K-SAT.
Notice that in all models 
\begin{equation}
F[0] = -\frac{1}{\beta} \langle \log ( 2 \cosh (\beta h )) \rangle_h |_{t=0}
\end{equation}

\subsubsection{$p$-spin}

In the case of the $p$-spin $H_J(S_{i_1},...,S_{i_p}) =
J\, S_{i_1}\cdot ...\cdot S_{i_p}$.
Substituting in eq.(\ref{pippo2}) and rearranging terms one
immediately finds:
\begin{eqnarray}
F_{var}^{p-spin}[G] 
 & = & \frac{1}{\beta} \left[ \alpha \left(p \left\langle \log (\cosh \beta u ) \right\rangle_u - \left\langle \log
 (\cosh \beta J ) \right\rangle_J \right) -
 \left\langle \log (2
 \cosh \beta h) \right\rangle_h + \right. \nonumber \\
& & \left. \alpha (p-1) \left\langle  \log \left(   1 + \tanh (\beta
 J)\prod_{t=1}^p \tanh (\beta g_t)    \right)   \right\rangle_{\{g_t\}, J}  \right]
\end{eqnarray}
while the remainder is the $t$ integral of 
\begin{eqnarray}
R_{RS}^{p-spin}[G,t]= & &
-\frac{\alpha}{\beta} \left[\frac{1}{N^p}
\sum_{i_1,...,i_p}
E \left\langle \log(1+\tanh(\beta J)\omega(S_{i_1}...S_{i_p})) \right\rangle_J 
-p E \left\langle \log(1+\tanh (\beta u)\omega
(S_i)) \right\rangle_u  + \nonumber \right. \\
& & \left. (p-1) E \left\langle 
\log(1+\tanh (\beta J)\prod_{t=1}^p \tanh (\beta g_p)) \right\rangle_{\{g_t\},J}
\right].
\end{eqnarray}
The expression for $F_{var}^{p-spin}[G]$ coincides with the RS free
energy once extremized over the variational space of probability
distributions $G$ \cite{tesi}. Terms have been properly added and
subtracted in order to get a remainder which equal to zero if
maximization over $G$ is taken, and the temperature is high enough for
replica symmetry to be exact \cite{tala-ksat}. 
As we will see, the remainder turns out to be
positive. $F_{var}^{p-spin}[G]$ is therefore, for all $G$ for which its
expression makes sense, a lower bound to the free-energy. At
saturation the condition 
\begin{equation}
G[h] = P[h] |_{t=0} \; \forall \; h
\end{equation}
should hold, which is simply the self-consistency RS equation.

By using equation
\begin{equation}
E \left\langle \log(1+\tanh (\beta u)\omega
(S_i)) \right\rangle_u = E \left\langle\log(1+\tanh (\beta J)
\prod_{t=1}^{p-1}\tanh (\beta g_r)
\omega (S_i)) \right\rangle_{\{g_t\},J}
\end{equation}
we can establish that the remainder is positive for even $p$.  We
expand the logarithm of the three terms in (absolutely converging)
series of $\tanh(\beta J)$, and notice that thanks to the parity of
the $J$ and the $g$ distributions, they will just involve negative
terms. We can then take the expected value of each terms and
write
\begin{equation}
R_{RS}^{p-spin}[G,t]= 
 \frac 1 \beta 
  \sum_{n=0}^{\infty} \langle \tanh^{2 n} \beta J \rangle_{J}
\frac{1}{n} \Omega \left[(q^{(2n)})^p - p q^{(2n)} 
\langle \tanh ^{2 n} \beta g \rangle_g^{p-1} + (p-1) \langle 
\tanh ^{2 n} \beta g
\rangle_g^p \right] 
\label{sser}
\end{equation}
where we have introduced the overlap $q^{(l)}$ and the replica measure
$\Omega$ defined in section 2. The series in (\ref{sser}) is an
average of positive terms in the case of the Viana-Bray model $p=2$,
where we get perfect squares, and more in general for all even $p$, as
we can easily, starting from the observation that in this case
$x^p-pxy^{p-1}+(p-1)y^p$ is positive or zero for all $x=q^{(2n)}$ ,$y=
\langle \tanh^{2 n} \beta J \rangle_{J}$ real.

In the case of $p$ odd, the same term is positive only if $x$ is
itself positive or zero. The bound of the free-energy would therefore
be established if we were able to prove that the probability
distributions of the $q^{(2n)}$ has support on the
positives.\footnote{A different sufficient condition for the series to
have positive terms is that $|\Omega(q^{(2n)})|\ge \langle \tanh(\beta
g)^{2n} \rangle_g$, but it is not clear its physical meaning.} This
property, which tells that anti-correlated states are not possible, is
physically very sound whenever the Hamiltonian is not symmetric under
change of sign of all spins. In fact, one expects the probability of
negative values of the overlaps to be exponentially small in the size
of the system for large $N$. Unfortunately however we have not been
able to prove this property in full generality. Notice that upon
maximization on $G$, the results of \cite{tala-ksat} imply that the
remainder is exactly equal to zero if the temperature is high enough
for replica symmetry to hold.

\subsubsection{K-SAT}

In the case of the $K$-SAT, using def.(\ref{SAT}) for the clause $H$, we find 
relation:
\begin{equation}
u_{\bf J}(h_1,...,h_{p-1}) \equiv u_J(\{J_t\},\{h_t\}) = \frac{J}{\beta} \tanh^{-1} \left[ 
\frac{\frac{\xi}{2} \prod_{t=1}^{p-1} \left(
\frac{1 + J_t \tanh( \beta h_t )}{2}  \right)}{1 +\frac{\xi}{2} \prod_{t=1}^{p-1} \left(
\frac{1 + J_t \tanh( \beta h_t )}{2}  \right) } \right] \; ,
\label{U-KSAT}
\end{equation}
where $\xi \equiv e^{-\beta} -1 < 0$.
Via direct inspection, the variational free-energy coincides with the
RS expression \cite{ksat}
\begin{eqnarray}
F_{var}^{K-SAT}[G] & = &
\frac{1}{\beta} \left[ \alpha (p-1) \left\langle \log   
\left( 1 + (e^{-\beta} -1)\prod_{t=1}^p \left( 
\frac{1 + \tanh (\beta g_t)}{2}  \right)  \right)  
\right\rangle_{\{g_t\},\{J_t\}}  - 
\langle  \log (2\cosh (\beta h)) \rangle_h + \right. \nonumber \\
& & \left. \alpha p \langle  \log (2\cosh (\beta u)) \rangle_u - \alpha p 
\left\langle \log   
\left( 1 + \frac{(e^{-\beta} -1)}{2}\prod_{t=1}^{p-1} \left( 
\frac{1 + \tanh (\beta g_t)}{2}  \right)  \right)  
\right\rangle_{\{g_t\},\{J_t\}} \right]
\label{FRSKSAT}
\end{eqnarray}
while the remainder is the $t$ integral of 
\begin{eqnarray}
R_{RS}^{K-SAT}[G,t] &=& -\frac{\alpha}{\beta} E \left[ 
\frac{1}{N^p} \sum_{i_1,...,i_p} \left\langle \log \left(  1+(\e^{-\beta}-1)\omega(\prod_{t=1}^p
	\frac{1+J_{t} S_{i_t}}{2}) \right)
\right\rangle_{\{J_t\}} 
- \right. \nonumber \\
& & \frac{p}{N} \sum_i \left\langle \log \left(  1+ \xi \omega\left(\frac{1+J
S_i}{2} \prod_{t=1}^{p-1}\frac{1+J_{t}
\tanh (\beta g_t)}{2} \right)  \right)
\right\rangle_{\{g_t\}, J, \{J_t\}} + \nonumber \\
& & \left. (p-1) \left\langle \log\left( 1+ \xi \prod_{t=1}^{p}\frac{1+J_{r}
\tanh (\beta g_t)}{2}\right) \right\rangle_{\{g_t\}, \{J_t\}} \right] \; .
\label{RESTOKSAT}   
\end{eqnarray}
Considerations analogous to the case of the $p$-spin, have led us to
add and subtract terms from eq.(\ref{pippo2}) to single out the proper
remainder term.  Expanding in series the logarithms, exploiting the
symmetry of the probabilities distribution functions and taking the
expectation of each term of the absolutely convergent series we
finally obtain:
\begin{equation}
R_{RS}^{K-SAT}[G,t] = \frac{\alpha}{\beta} \sum_{n\ge 1} \frac{(-1)^{n}}{n} (\xi^*)^n
\Omega \left[
(1+Q_n)^p - p(1+Q_n)\langle(1+J \tanh (\beta g))^n\rangle_{J,g}^{p-1}
+(p-1) \langle(1+J \tanh (\beta g))^n\rangle_{J,g}^{p} \right]
\label{RESTO2KSAT}
\end{equation}
where we have defined $\xi^* \equiv \xi/(2^p) < 0$ and $Q_n \equiv
\sum_{l=1}^n \langle J^l \rangle_J \sum_{a_1<...<a_l}^{1,n}
q^{a_1...a_l}$.  Detailed calculations are given in the appendix.  As
in the $p$-spin case, the previous sum is obviously positive for $p$
even. For $p$ odd we should again rely on the physical wisdom that
all $q^{(a_1,...,a_l)}$ have positive support and so have the
functions $1+Q_n \ge 0$.  Again, the variational free-energy coincides
with the RS expression once extremized over $G$ at the condition $P =
G$ at $t=0$.

\subsection{The 1RSB Bound}
\label{tarta}

We establish here a more complex estimate, in a larger
variational space of functional probability distributions. 
The general strategy will be here to consider the same form 
for the auxiliary
Hamiltonian, but now with a more involved choice for
the fields distribution. The fields on different sites or different
index $l_i$ will be
still independent, but each site field distribution
$G_i^{l_i}(g_i^{l_i})$ will be itself random i.i.d., 
chosen with a probability density functional ${\cal G}[G]$, with support
on symmetric distributions $G(-g)=G(g)$. It will be assumed that
${\cal G}$ is such that all the expressions below make sense. 
In this case, the variational approximation for the free-energy will be obtained from an
estimate of 
\begin{equation}
-\beta F_N[m,t] = \frac{1}{mN} E_1 \log E_2(Z^m(t))
\end{equation}
where we have denoted with:
\begin{itemize}
\item  $E_2$ the average w.r.t. $g_i^{l_i,n}$ for
fixed distributions $G_i^{l_i,n}$ according to the measure
\begin{equation}
C \prod_{i=1}^N\prod_{l_i=1}^{k_i} \prod_{n=1}^{p-1} dg_i^{l_i,n} \;
G_i^{l_i,n}(g_i^{l_i,n})
\left(\frac{B_{J_i^{l_i,n}}(g_i^{l_i,1}...g_i^{l_i,p-1})}{2\cosh(\beta
u_{J_i^{l_i,n}}(g_i^{l_i,1}...g_i^{l_i,p-1})
)}\right)^m
\end{equation}
where $C$ ensures the normalization. 

\item  $E_1$ the average with respect
to the quenched clause variable, 
distributions the $G_i^{l_i}$'s and the Poissonian variables $k_i$'s,
which will be i.i.d. with probabilities $\mu(J)$, 
${\cal G}(  G_i^{l_i})$ and $\pi(k_i,(1-t)\alpha)$ respectively. 
\end{itemize}
The number $m$ is real in the interval (0,1]. 
The statistical ensemble of the 
auxiliary fields $u$ and $h$ will be now related to the one of the $g$
by:
\begin{equation}
{\cal Q}(Q)=\int {\cal D}G_1 {\cal G}(G_1) ... {\cal D}G_{p-1} {\cal
G}(G_{p-1}) \langle \delta (Q(\cdot)-
Q(\cdot|G_1,...,G_{p-1},J))\rangle_J
\label{probQ2}
\end{equation}
\begin{equation}
{\cal P}(P) = \sum_{k=0}^\infty \e^{-\alpha p (1-t)}\frac {(\alpha p (1-t))^k}{k!} 
\int \prod_{l=1}^{k} {\cal D}{Q_l}
{\cal Q}({Q_l})
\delta(P(\cdot)- P(\cdot|{Q_1},...,{Q_k} ) )
\label{ProbP2}
\end{equation}
 where:
\begin{eqnarray}
& & Q(u|G_1,...,G_{p-1},J) =  {\cal N}_G [G_1,...,G_{p-1}]\int dg_1\; G_1(g_1) 
...dg_{p-1}\; G_1(g_{p-1}) B_J(g_1,...,g_{p-1})^m 
\delta (u-u_J(g_1,...,g_{p-1}) ) \\
& & G(g|{Q_1},...,{Q_k} ) = {\cal N}_{Q,k}[{Q_1},...,{Q_k}] \, \,
(2\cosh(\beta g))^m \int \prod_{l=1}^k
d u_l \; \frac{ {Q_l} (u_l)}{(2\cosh(\beta u_l))^m} 
\delta ( g - \sum_{l=1}^k u_l )
\label{innerprobs}
\end{eqnarray}
where ${\cal N}_{Q,k}[{Q_1},...,{Q_k} ]$, ${\cal N}_G
[G_1,...,G_{p-1}]$ and $B_J(g_1,...,g_{p-1})$ have been previously
defined.
With notations similar to the ones of the RS case, we can write 
\begin{eqnarray}
\frac{d}{dt} (-\beta F_N[m,t])= & & -\alpha \sum_M (\pi(M,\alpha t
N)-\pi(M-1,\alpha t N)) E'_1 \frac {1}{Nm} \log E_2 Z(t)^m\nonumber
 \\
& &+\frac{\alpha p}{N} \sum_i \sum_{k_i}
(\pi(k_i,\alpha p(1-t))-\pi(k_i-1,\alpha p(1-t))) E''_{1,i} \frac {1}{Nm} \log E_2 Z(t)^m
\label{der-rsb}
\end{eqnarray}
which, extracting explicitly the contribution from the $M$-th close in the 
first term and the $k_i$-th field $u$ in the second, following
considerations similar to the RS case we find:
\begin{eqnarray}
\frac{d}{dt} (-\beta F_N[m,t])= & & \alpha \sum_M (\pi(M-1,\alpha t N)) \frac{1}{m}E'_1
\log \left[
\frac
{E_2 Z_{-M}^m \omega(\e^{-\beta H_{J^{(\mu)}}(S_{i_1^M},...,S_{i_p^M})})_{-M}^m}
{E_2 Z_{-M}^m}
\right]\nonumber
\\& & -\frac{p
\alpha}{N} \sum_i\sum_{k_i} \pi(k_i-1,\alpha p(1-t))
\frac{1}{m}E''_{1,i}
\log \left[
\frac
{E_2 Z_{-u_i^{k_i}}^m  \omega(\e^{\beta
u_i^{k_i}S_i})_{-u_i^{k_i}}^m}
{E_2 Z_{-u_i^{k_i}}^m }
\right]. 
\label{pippo3}
\end{eqnarray}
Again it can be recognized that the primed averages coincide with the
averages over the original ensembles plus the averages on the
variables appearing in the terms we extracted. Finally we get:
\begin{equation}
\frac{d}{dt} (-\beta F_N[m,t])=\frac{\alpha}{m} E_1 \left[
\frac{1}{N^p}\sum_{i_1,...,i_p} \left\langle 
\log \left(
\frac{E_2 Z^m \omega (e^{-\beta H_J(S_{i_1},...,S_{i_p})})^m}{E_2 Z^m}
\right)
\right\rangle_J - \frac{p}{N}\sum_i \left\langle 
\log \left(
\frac{E_2 Z^m \langle \omega (e^{\beta u S_i})^m \rangle_u}{E_2 Z^m}
\right)
\right\rangle_Q
\right] \; . 
\label{pippo4}
\end{equation}
Rearranging all terms
one finds the estimate: 
\begin{equation}
F_N = F_{var}[{\cal G}] + \int_0^1 dt\; R_{1RSB}[{\cal G},t]+O(1/N)
\end{equation} 
where this time $ F_{var}[{\cal G}]$ coincides with $F_{1RSB}[{\cal G}]$, 
the expression of the variational free-energy in
the 1RSB treatment at the saddle point ${\cal G} = {\cal P}$ at $t=0$,
and $\int_0^1 dt\; R_{1RSB}[{\cal G},t]$ is the remainder. 
Notice that the derivation immediately suggests how to 
generalize the analysis to more steps of replica symmetry breaking. 
Let us now specialize the formulae for the 
$p$-spin model and the K-SAT.
Again, in this case we will need the expression for $F[0]$:
\begin{equation}
F[0] = \frac{1}{\beta m} \left[
\left\langle  \log \left\langle  \left( \frac{1}{2 \cosh (\beta h)} \right)^m \right\rangle_h
\right\rangle_P \right]_{|_{t=0}} \; .
\end{equation} 

\subsubsection{$p$-spin}
\label{sec-p-spin1RSB}

In this case, plugging def.(\ref{spin}) in eq.(\ref{pippo4})
rearranging, adding and subtracting terms one finds: 
\begin{eqnarray}
F_{var}^{p-spin}[{\cal G}] &=& \frac{1}{\beta m} \left[
\left\langle  \log \left\langle  \left( \frac{1}{2 \cosh (\beta h)} \right)^m \right\rangle_h
\right\rangle_P - \alpha m  \left\langle \log (2 \cosh (\beta J))
\right\rangle_J - \alpha p \left\langle \log \left\langle  \left( \frac{1}{2 \cosh
(\beta u)}\right)^m    \right\rangle_u   \right\rangle_Q
\right. \nonumber \\ & & \left. + \alpha
(p-1)  \left\langle \log \left\langle  \left(  1 + \tanh (\beta J)
\tanh (\beta g_1)... \tanh (\beta g_p) \right)^m
\right\rangle_{g_1,...,g_p} \right\rangle_{G_1,...,G_p; J} \right] 
\label{1RSB-variational-spin}
\end{eqnarray}
while the remainder is the $t$ integral of 
\begin{eqnarray}
R_{1RSB}^{p-spin}[{\cal G},t]= & &
-\frac{\alpha}{\beta m} E_1 \left[ \frac{1}{N^p} \sum_{i_1,...,i_p}
\left\langle 
\log \left(
\frac{E_2 Z^m (1 + \omega (S_{i_1}...S_{i_p})\tanh (\beta J))^m}{E_2 Z^m}
\right)
\right\rangle_J - \right. \nonumber \\
& & 
\frac{p}{N} \sum_i \left\langle
\log \left(
\frac{E_2 Z^m \left\langle (1+ \omega(S_i)\tanh (\beta u ))^m \right\rangle_u}{E_2 Z^m}
\right)
\right\rangle_Q
+ \nonumber \\
& & \left. (p-1) \left\langle \log \left\langle  \left(  1 + \tanh (\beta J)
\tanh (\beta g_1)... \tanh (\beta g_p) \right)^m
\right\rangle_{g_1,...,g_p} \right\rangle_{G_1,...,G_p; J} 
\right]
\label{pippospin}
\end{eqnarray}
The expression for $F_{var}^{p-spin}[{\cal G}]$ coincides with the
1RSB free-energy \cite{tesi}
once maximized over the variational space of
probability distribution functionals ${\cal G}$. The maximization
condition reads:
\begin{equation}
{\cal G}[P] = {\cal P}[P] \; |_{t=0} \; \forall \; P \; ,
\label{SPERSB}
\end{equation}
which is simply the self consistency 1RSB condition.
For even $p$ (and in particular for $p=2$ that corresponds to the
Viana-Bray case), one can check that the remainder is positive just
expanding the logarithm in series and exploiting the parity of the $J$
and the $g$ distributions. As this is considerably more involved then
in the RS case, we relegate this check to appendix A.

\subsubsection{$K$-SAT}
\label{sec-K-SAT1RSB}

In the $K$-SAT case the expression for function $B_{\bf
J}(h_1,...,h_{p-1})$ reads:
\begin{equation}
B_{\bf J}(h_1,...,h_{p-1}) \equiv B(\{J_t\},\{h_t\}) = 1 + \frac{\xi}{2} \prod_{t=1}^{p-1} \left(
\frac{1 + J_t \tanh( \beta h_t )}{2}  \right) \; ,
\label{B-u-KSAT}   
\end{equation}
while the corresponding one for $u_{\bf J}(h_1,...,h_{p-1})$ is the same as in the RS case.
The corresponding replica free-energy and remainder read
\begin{eqnarray}
F_{var}^{K-SAT}[{\cal G}] &=& \frac{1}{m\beta} \left[ \alpha (p-1)
\left\langle  \log \left\langle    
\left( 1 + \xi \prod_{t=1}^p \left( \frac{1 + J_t \tanh (\beta g_t)}{2}   \right)  \right)^m
\right\rangle_{\{ g_t\}}  
\right\rangle_{\{G_t\},\{J_t\}} \right. - \nonumber \\
& & \left. \alpha p \left\langle \log \left\langle 
\left( \frac{B(\{J_t\},\{g_t\})}{2 \cosh (\beta u_J(\{J_t\},\{g_t\})  )}   \right)^m
\right\rangle_{\{g_t\}}  \right\rangle_{\{G_t\},\{J_t\},J} + 
\left\langle \log \left\langle \left( \frac{1}{2 \cosh (\beta h)}  \right)^m \right\rangle_h
\right\rangle_P \right]
\label{liuto}
\end{eqnarray}
The remainder is the $t$ integral of 
\begin{eqnarray}
R_{1RSB}^{K-SAT}[{\cal G},t] & = & -\frac{\alpha}{\beta m} E_1 \left[
\frac{1}{N^p} \sum_{i_1,...,i_p}
\left\langle 
\log \left(
\frac{E_2 Z^m \left( 1 + \xi \omega \left( \prod_{t=1}^p \frac{1 + J_t
S_{i_t}}{2}  \right)  \right)^m}{E_2 Z^m}
\right) \right\rangle_{\{J_t\}} \right. - \nonumber \\
& & \frac{p}{N} \sum_i \left\langle
\log \left(
\frac{E_2 Z^m \left\langle    
\left(
1 + \xi \frac{1 + J \omega (S_i)}{2} \prod_{t=1}^{p-1} \frac{1 + J_t
\tanh (\beta g_t)}{2} 
\right)^m
\right\rangle_{\{ g_t\}}}{E_2 Z^m}
\right)
\right\rangle_{\{ G_t\},\{J_t\},J} + \nonumber \\
 & & \left. (p-1) 
\left\langle  \log \left\langle    
\left( 1 + \xi \prod_{t=1}^p \left( \frac{1 + J_t \tanh (\beta g_t)}{2}   \right)  \right)^m
\right\rangle_{\{ g_t\}}  
\right\rangle_{\{G_t\},\{J_t\}} \right]
\label{pipposat}
\end{eqnarray}
The expression for $F_{var}^{K-SAT}[{\cal G}]$ coincides with the 1RSB free
energy once extremized under condition (\ref{SPERSB}),
with the corresponding $K$-SAT probability distribution
functionals. Notice that
The proof of the positivity of (\ref{pipposat}) for even $p$ is again dove via
series expansion, all the detail are explained in Appendix B. 

At this point we can take the zero temperature limit, finding that the
resulting expression gives us a lower bound for the ground-state
energy of the system, i.e. the minimal number of unsatisfied
clauses. Notice that the $T \to 0$ limit of the replica free-energy is
not trivial. The necessary assumptions on the field distributions to
get it correct are well known in the physical literature, and have
been recently reviewed in \cite{MP}.  Recently Mezard, Parisi and
Zecchina \cite{MPZ} have worked out the K-SAT 1RSB solution for
$p=3$ predicting a non zero ground-state energy for values of $\alpha$
above a satisfiability threshold of $\alpha_c=4.256$, very well in
agreement with the numerical simulations. Our results, together with
the additional hypothesis of positivity of the support of the overlap
functions imply that this value is an upper bound to the true
threshold.

\section{Summary and conclusions}
In this paper we have established that the free-energy of some
families of diluted random spin models can be written as the sum of a
term identical to the ones got in the cavity/replica plus an error
term. Both the replica term and the remainder are different in
different replica scheme, corresponding to the choice of statistical
ensemble of the cavity fields. We believe that the sign of the
remainder is in general negative in the model we have considered,
although we have been able to prove that only in the case of 
even $p$. For odd $p$ our belief is supported by the physical
wisdom that the overlap distributions are supported on the
positives in the large $N$ limit. 

We have considered the cases of replica symmetry and one step of
replica symmetry breaking. It is clear that the analysis could be
extended to further levels of replica symmetry breaking, although the
complexity of the analysis would greatly increase. The 1RSB level is
thought to give the exact scheme to treat the $p$ spin model and the
K-SAT problem for $p\ge 3$. For the Viana-Bray model on the other hand
it is believed that no finite RSB scheme furnish the exact solution,
and one needs to consider the limit of infinite number of replica
symmetry breaking. It is not clear to us how to generalize the
analysis to this case.

Our analysis of the diluted models underlines a strong link between
the Guerra method and the cavity method which remained rather hidden
in the fully connected case. In the cavity approach one considers
incomplete graphs in which either sites or clauses are removed from
the complete graph. Then, with the aid of precise physical hypothesis,
consistency equations are written that allow to compute the
free-energy from the comparison between the site and clause
contributions. In the approach presented in this paper the removal of
clauses is compensated in average by the addition of some external
fields which have precisely the statistics which is assumed with
cavity. The novelty of the approach is that it gives some control on
the approximation involved, and proves the variational nature of the
replica free-energies. Of course a complete control on the remainder
in various situations would result in rigorous solutions.

Although in this paper we have mainly worked at finite temperature,
the zero temperature limit can be considered without harm. This is
particularly relevant in random satisfiability problem, where it is
typically found a SAT-UNSAT transition where the ground state energy
passes from zero to non zero values. Our analysis shows that, the
replica estimates for many of the models considered in the literature,
and possibly some of the ones to be obtained in the future with the
same method provide upper bounds for the satisfiability thresholds.

In this paper we have confined ourselves to spin models on graphs with
Poissonian connectivity. The extension to more general diluted graph
models will be presented in a forthcoming paper. 

Finally we would like to remark that despite the heavy formalism, our
proofs to the bounds are conceptually simple. They are issue of
explicit computations and elementary positivity consideration. We hope
that this contributes to illustrate the elegance of the construction
first introduced in \cite{guerra}.

\section{Acknowledgments}
We thank M. Isopi for pointing us out the importance of
ref. \cite{guerra}. We thank G. Gaeta for discussions.  

\appendix

\section{p-spin}

\subsection{Check of the positive sign of $R_{1RSB}^{p-spin}$}
\label{spinRSB}

In this appendix we will explicitly show that expression
(\ref{pippospin}) has positive sign. 
We proceed
expanding in absolutely convergent series each of the three addend in 
(\ref{pippospin}) and showing, 
taking the expectation 
of each term that the resulting series is positive semidefinite.

The first term writes:  
\begin{eqnarray}
&  
E_1 \left\langle \log \frac{E_2 Z^m 
(1+\tanh (\beta J) \omega (S_{i_1}... S_{i_p}))^m}
{ E_2 Z^m} \right\rangle_J = &\nonumber \\
& \sum_{l\ge 1}\frac{(-1)^{l+1}}{l}\sum_{k_1,...,k_l}^{1,\infty}
\left(\frac{m (-1)^{k_1-1}}{k_1!} \prod_{r_1=1}^{k_1-1}(r_1-m)\right)...
\left(\frac{m (-1)^{k_l-1}}{k_l!} \prod_{r_l=1}^{k_l-1}(r_l-m)\right)
\left\langle (\tanh(\beta J)^{\sum_{s=1}^l k_s} \right\rangle_J \cdot
& \nonumber \\
& E_1\left( \prod_{s=1}^l \frac {E_2(Z^m \omega(S_{i_1}... S_{i_p})^{k_s})}
{E_2(Z^m)}\right) &
\label{medusa}
\end{eqnarray}
where the term $E_1 (\, \cdot \,)$ in the last line of eq.(\ref{medusa}) 
can be written as 
\begin{equation}
E_1\left( \frac{E_2^{(1)}...E_2^{(l)} Z_{(1)}^m...Z_{(l)}^m
\omega_{(1)} (S_{i_1}^{1,1}... S_{i_p}^{1,1} ... S_{i_1}^{k_1,1}
... S_{i_p}^{k_1,1})...  \omega_{(l)} (S_{i_1}^{1,l}... S_{i_p}^{1,l}
... S_{i_1}^{k_l,l} ... S_{i_p}^{k_l,l}) }{(E_2 Z^m)^l}
\right)=\Omega^{(l)} \left[ (q^{(k_1,...,k_l)})^p \right] \; .
\end{equation}
where each $\omega_{(s)}$ ($s=1,...,l$) is a product of $k_s$ Gibbs
measure with independent fields (variables appearing in the
$E_2^{(s)}$ averages), and same fields distributions and quenched disorder
(variables appearing in $E_1$). 
The quantities $q^{(k_1,...,k_l)}$ have been defined as:
\begin{equation}
q^{(k_1,...,k_l)} =  \frac{1}{N}\sum_i S_i^{1,1} \cdot ...\, \,
\cdot \, S_i^{k_1,1}  \cdot ...
\, \, \cdot \, S_i^{1,l}  \cdot ... \, \, \cdot \, S_i^{k_l,l}
\label{QK}
\end{equation}
and in this case the averages are performed using a 
a generalized replica measure, defined as:
\begin{equation}
\Omega^{(l)} [(q^{(k_1,...,k_l)})^n] = 
E_1 \left[ \frac{\prod_{s=1}^l E_2^{(s)}  Z_{(s)}^m \omega_{(s)} (S_{i_1}...S_{i_n})^{k_s}}
{(E_2 Z^m)^l} \right]
\label{gen-repmeas}
\end{equation}
for any integer $n$. 
The average over $J$ selects the terms with
even $\sum_{s=1}^l k_l$  in (\ref{medusa}) so that we finally find 
\begin{equation}
-\sum_{l\ge 1}\frac{m^l}{l}\sum_{{k_1,...,k_l \atop \sum_{s=1}^l k_s {\rm
 even}}}^{1,\infty} \prod_{s=1}^l \left(
 \frac{\prod_{r_s=1}^{k_s-1}(r_s-m)}{k_s!} \right) \left\langle
 (\tanh(\beta J))^{\sum_{s=1}^l k_s} \right\rangle_J \Omega^{(l)} \left[ (q^{(k_1,...,k_l)})^p \right]
\end{equation}
notice that $(r_s-m) \ge 0$ $\forall$ integer $r_s > 0$ only in the
current hypothesis that $m \in [0,1]$. 
Analogously, the term 
\begin{equation}
E_1 \left\langle \log
\frac{
E_2 Z^m \left\langle (1+\tanh(\beta u)\omega(S_i))^m \right\rangle_u}
{E_2 Z^m} \right\rangle_Q
\end{equation}
writes
\begin{equation}
-\sum_{l\ge 1}\frac{m^l}{l}\sum_{{k_1,...,k_l \atop \sum_{s=1}^l k_s
 {\rm even}}}^{1,\infty} \prod_{s=1}^l \left(
 \frac{\prod_{r=1}^{k_s-1}(r-m)}{k_s!} \right) \left\langle
 \prod_{s=1}^l \left\langle \tanh(\beta u)^{k_s} \right\rangle_u
 \right\rangle_Q \Omega^{(l)} \left[ (q^{(k_1,...,k_l)}) \right]
\end{equation}
or, making use of the definition of $G(g)$,
\begin{equation}
-\sum_{l\ge 1}\frac{1}{l}\sum_{{k_1,...,k_l \atop \sum_{s=1}^l k_s {\rm
 even}}}^{1,\infty} \prod_{s=1}^l \left(
 \frac{\prod_{r=1}^{k_s-1}(r-m)}{k_s!} \right) \left\langle
 \prod_{s=1}^l \left\langle  (\tanh(\beta g))^{k_s} \right\rangle_g
 \right\rangle_G^{p-1} \left\langle (\tanh(\beta J))^{\sum_{s=1}^l k_s}
 \right\rangle_J \Omega^{(l)} \left[ (q^{(k_1,...,k_l)}) \right]
\end{equation}
Eventually, following analogous manipulations, the last term
\begin{equation}
\left\langle \log \left\langle \left( 1+\tanh(\beta J)\prod_{t=1}^{p}
\tanh(\beta g_t) \right)^m
\right\rangle_{\{g_t\}} \right\rangle_{J,\{ G_t  \}}
\end{equation}
can be written as 
\begin{equation} 
-\sum_{l\ge
 1}\frac{m^l}{l}\sum_{{k_1,...,k_l \atop \sum_{s=1}^l k_s {\rm
 even}}}^{1,\infty} \prod_{s=1}^l \left(
 \frac{\prod_{r=1}^{k_s-1}(r-m)}{k_s!} \right) \left\langle
\prod_{s=1}^l \left\langle (\tanh(\beta g))^{k_s} \right\rangle_g 
\right\rangle_G^p
 \left\langle (\tanh(\beta J))^{\sum_{s=1}^l k_s} \right\rangle_J .
\end{equation}
Invoking \ref{probQ2} and collecting all
\begin{eqnarray} 
R_{1RSB}^{p-spin}[{\cal G},t] & = & \frac{\alpha}{\beta m} 
\sum_{l\ge 1}\frac{m^l}{l} 
\sum_{{k_1,...,k_l \atop \sum_{s=1}^l k_s {\rm
 even}}}^{1,\infty} 
\prod_{s=1}^l 
\left(
 \frac{\prod_{r=1}^{k_s-1}(r-m)}{k_s!} 
\right) 
\left\langle 
\left( 
\tanh (\beta J) 
\right)^{\sum_{s=1}^l k_s} 
\right\rangle_J 
\; \cdot \nonumber \\
& &  \Omega^{(l)} \left[
(q^{(k_1,...,k_l)})^p -
p A(k_1,...,k_l)^{p-1}  
(q^{(k_1,...,k_l)}) + (p-1) A(k_1,...,k_l)l^{p}  
\right]
\label{final_monster}
\end{eqnarray}
where we have defined:
\begin{equation}
A(k_1,...,k_l) \equiv  \left\langle \prod_{s=1}^l \left\langle (\tanh (\beta g))^{k_s} 
\right\rangle_g
\right\rangle_G
\label{AL} 
\end{equation}
Each inner term of the series (\ref{final_monster})
\begin{equation}
\Omega^{(l)} \left[
(q^{(k_1,...,k_l)})^p -
p A(k_1,...,k_l)^{p-1}  
(q^{(k_1,...,k_l)}) + (p-1) A(k_1,...,k_l)^{p}  
\right]
\label{final_term}
\end{equation}
is always positive semidefinite for $p$ even while we need the
condition conditions $q^{(k_1,...,k_l)} \ge 0$ for $p$ odd.  For $p=2$
one retrieves the Viana-Bray result where (\ref{final_term}) is a
perfect square.  As in the RS case, one can now integrate
eq.(\ref{final_monster}) and recognize that once more the total true
free-energy can be written as variational term plus a positive extra
one. The variational term coincides with the 1RSB free-energy at
stationarity and under condition
\begin{eqnarray}
{\cal G} (P) &=& {\cal P} (P)|_{t=0}  \, \, \, \, \,  \forall \, \, \, P \; .
\end{eqnarray}

\section{K-SAT}

\subsection{Check of the positive sign of $R_{RS}^{K-SAT}$ ...}
 
The aim of this appendix is to show that the expression for the
remainder $R_{RS}[G,t]$ in (\ref{FveraRS}) for the $K$-SAT model
case as positive sign. 
For the $K$-SAT $R_{RS}[G,t]$ specializes to:\footnote{The sum of the
site indices has been eliminated by symmetry.}
\begin{eqnarray}
R_{RS}^{K-SAT}[G,t] &=& -\frac{\alpha}{\beta} E \left[ \left\langle \log \left( \omega 
\left( \exp^{-\beta \prod_{r=1}^p \frac{1+J_{r} S_{r}}{2}} \right) \right)
\right\rangle_{\{J_t\}} -  \right. \nonumber \\ & & p \left\langle
\log \left(    
1 + \omega(S) \tanh (\beta u) 
\right)
\right\rangle_u  - 
 - p \left\langle \log \left(  1 + \frac{\xi}{2}
\prod_{t=1}^{p-1} \left( \frac{1 + J_t \tanh (\beta g_t)}{2} \right)  \right)
\right\rangle_{\{g_t\}, \{J_t\}} + \nonumber \\
& & \left. (p-1) \left\langle \log\left( 1+ \xi \prod_{t=1}^{p}\frac{1+J_{r}
\tanh (\beta g_t)}{2}\right) \right\rangle_{\{g_t\}, \{J_t\}} \right] 
\label{REM-KSAT-RS-1}
\end{eqnarray}
which thanks to the relation between $Q(u)$ and $G(g)$, rewrites as
\begin{eqnarray}
R_{RS}^{K-SAT}[G,t] &=& -\frac{\alpha}{\beta} E \left[ \left\langle \log \left(  1+(\e^{-\beta}-1)\omega(\prod_{t=1}^p
	\frac{1+J_{t} S_{t}}{2}) \right)
\right\rangle_{\{J_t\}} 
- \right. \nonumber \\
& & p \left\langle \log \left(  1+ \xi \omega\left(\frac{1+J
S}{2} \prod_{t=1}^{p-1}\frac{1+J_{t}
\tanh (\beta g_t)}{2} \right)  \right)
\right\rangle_{\{g_t\}, J, \{J_t\}} + \nonumber \\
& & \left. (p-1) \left\langle \log\left( 1+ \xi \prod_{t=1}^{p}\frac{1+J_{r}
\tanh (\beta g_t)}{2}\right) \right\rangle_{\{g_t\}, \{J_t\}} \right] 
\label{REM-KSAT-RS}
\end{eqnarray}
The last term has been added and subtracted from
eq.(\ref{FveraRS})
in order to extract a remainder that would vanish if replica symmetry
holds, and maximization is performed on $G(g)$. 
As in the $p$-spin case, we will proceed in a Taylor expansion
of expression (\ref{REM-KSAT-RS}) in powers of $\xi$, and rely on
absolute convergence to average each term of the series.  

Expanding the first term in (\ref{REM-KSAT-RS}) we can write 
\begin{eqnarray}
& &E \left[ \left\langle \log\left(1+\xi\omega(
	\prod_{t=1}^{p}
	\frac{1+J_{t} S_{t}}{2})\right) \right\rangle_{\{J_t\}}\right]
	=
\nonumber
\\
& &
\sum_{n\ge 1} \frac{(-1)^{n+1}}{n} (\xi^*)^n E \left[ \left\langle 
\omega\left(\prod_{t=1}^p 
(1+J_t S_t)\right)^n \right\rangle_{\{J_t\}} \right] =\nonumber
\\
& &
\sum_{n\ge 1} 
\frac{(-1)^{n+1}}{n} 
(\xi^*)^n 
\Omega \left[ \prod_{t=1}^p \left( 1+ \sum_{l=1}^n \left\langle J^l_t \right\rangle_{J_t}
\sum_{a_1<...<a_l}^{1,n} S_t^{a_1}...S_t^{a_l}\right)
      \right]=\nonumber
\\
& &
\sum_{n\ge 1}
\frac{(-1)^{n+1}}{n} 
(\xi^*)^n 
\Omega \left[
	\prod_{t=1}^p 
	\left( 1+ 
		\sum_{l=1}^n \left\langle J^l_t \right\rangle_{J_t} \sum_{a_1<...<a_l}^{1,n} 
		q^{a_1...a_l}
	\right)
      \right]=\nonumber
\\
& & \sum_{n\ge 1} \frac{(-1)^{n+1}}{n} (\xi^*)^n 
\Omega[(1+Q_n)^p]
\end{eqnarray}
where we have defined $\xi^* \equiv  (e^{-\beta} - 1)/(2^p) $ and 
$\sum_{l=1}^n  \left\langle J^l \right\rangle_J
\sum_{a_1<...<a_l}^{1,n} 
q^{a_1...a_l} \equiv Q_n$. Notice that due to the negative sign of
$\xi^*$, the coefficients $(-1)^{n+1}(\xi^*)^n$ are all negative. 

The analogous expansion of the second term is:
\begin{eqnarray}
& & E \left[\left\langle \log\left( 1+ \xi \omega\left(\frac{1+J
S}{2} \prod_{t=1}^{p-1}\frac{1+J_{t} \tanh (\beta g_t)}{2} \right)\right) \right\rangle_{\{J_t\}, J, \{g_t\}}\right]=
\nonumber
\\
& & 
\sum_{n\ge 1} \frac{(-1)^{n+1}}{n} (\xi^*)^n \Omega \left[ 
\left( 1+ \sum_{l=1}^n \langle J^l \rangle_J \sum_{a_1<...<a_l}^{1,n}
q^{a_1...a_l} 
\right) 
\left\langle
\prod_{t=1}^{p-1} \prod_{l=1}^{n} \left( 1 + J_t \tanh (\beta g_t )  \right)
\right\rangle_{\{J_t\},\{g_t\}} \right] = \nonumber
\\
& &
\sum_{n\ge 1} \frac{(-1)^{n+1}}{n} (\xi^*)^n\Omega \left[ (1+Q_n)\left\langle(1+J
\tanh (\beta g))^n\right\rangle_{J,g}^{p-1} \right]
\end{eqnarray}
Finally, the third terms in eq.(\ref{REM-KSAT-RS}) immediately reads
\begin{equation}
\left\langle \log\left( 1+ \xi \prod_{t=1}^{p}\frac{1+J_{t}
\tanh (\beta g_t)}{2}\right) \right\rangle_{\{J_t\}, \{g_t\}}=
\nonumber
\\
\sum_{n\ge 1} \frac{(-1)^{n+1}}{n} (\xi^*)^n \langle(1+J \tanh (\beta g))^n\rangle_{J,g}^{p}
\end{equation}
The sum of the three pieces in eq.(\ref{REM-KSAT-RS}) gives:
\begin{equation}
R_{RS}^{K-SAT}[G,t] = \frac{\alpha}{\beta} \sum_{n\ge 1} \frac{(-1)^{n}}{n} (\xi^*)^n
\Omega \left[
(1+Q_n)^p - p(1+Q_n)\langle(1+J \tanh (\beta g))^n\rangle_{J,g}^{p-1}
+(p-1) \langle(1+J \tanh (\beta g))^n\rangle_{J,g}^{p} \right]
\end{equation}
The previous sum is always positive semidefinite for $p$ even while we
need $ 1+Q_n \ge 0$ for $p$ odd.

\subsection{...and of $R_{1RSB}^{K-SAT}$}
\label{SATRSB}

We proceed in the same way as in the $p$-spin case. The algebra is
elementary but more tedious and involved, therefore we will only list
the final results of the calculation.
Starting from eq.(\ref{pipposat}), we again expand in series the first term,
getting, with a treatment similar to the RS case:
\begin{eqnarray}
& & 
R_{1RSB}^{K-SAT}[{\cal G},t] = \sum_{l\ge 1}\frac{m^l}{l} 
\sum_{k_1,...,k_l}^{1,\infty} (-\xi^*)^{\sum_{s=1}^l k_s} \prod_{s=1}^l 
\left(
 \frac{\prod_{r=1}^{k_s-1}(r-m)}{k_s!} 
\right) 
\Omega^{(l)}
\left[
(1 + {\bf Q}(k_1,...,k_l))^p
\right]
\label{primomostro}
\end{eqnarray}
where we have defined:
\begin{eqnarray}
{\bf Q}(k_1,...,k_l) &\equiv & \sum_{s=1}^l \sum_{r_1,...,r_s}^{k_1,...,k_s}
\left\langle
J^{(r_1 + ... + r_s)}
\right\rangle_J \prod_{t=1}^s
\sum_{a_1 < ... < a_{r_t} = 1}^{k_1,...,k_s}
q^{(a_{r_1},...,a_{r_s})} 
\label{secondomostro}
\end{eqnarray}
Analogous steps give for the second term in eq.(\ref{pipposat})
\begin{eqnarray}
& & \sum_{l\ge 1}\frac{m^l}{l} 
\sum_{k_1,...,k_l}^{1,\infty} (-\xi^*)^{\sum_{s=1}^l k_s}
\prod_{s=1}^l 
\left(
 \frac{\prod_{r=1}^{k_s-1}(r-m)}{k_s!} 
\right)
 \Omega^{(l)}
\left[
1 + {\bf Q}(k_1,...,k_l)
\right]
 \left\langle \prod_{s=1}^l \left\langle \left(
1 + J \tanh (\beta g )
\right)^{k_l}
\right\rangle_g   \right\rangle_{G,J}^{p-1}
\label{terzomostro}
\end{eqnarray}
and for the third term
\begin{eqnarray}
& &  \sum_{l\ge 1}\frac{m^l}{l} 
\sum_{k_1,...,k_l}^{1,\infty} 
(-\xi^*)^{\sum_{s=1}^l k_s}
\prod_{s=1}^l 
\left(
 \frac{\prod_{r=1}^{k_s-1}(r-m)}{k_s!} 
\right)  \left\langle \prod_{s=1}^l \left\langle \left(
1 + J \tanh (\beta g )
\right)^{k_l}
\right\rangle_g   \right\rangle_{G,J}^p \; ,
\label{quartomostro} 
\end{eqnarray}
where in the last two terms we can further expand
\begin{equation}
\left\langle \prod_{s=1}^l \left\langle \left(
1 + J \tanh (\beta g )
\right)^{k_l}
\right\rangle_g   \right\rangle_{G,J}^n =
\left(
\sum_{r_1,...,r_l = 1}^{k_1,...,k_l}
\prod_{s=1}^l 
\left(k_s \atop{r_s}\right)
\left\langle J^{(r_1 + ... + r_l)}
\right\rangle_J
\left\langle
\prod_{s=1}^l
\left\langle
\left(
\tanh (\beta g)
\right)^{r_s}
\right\rangle_g
\right\rangle_G^n
\right) 
\label{quintomostro}
\end{equation}
with $n$ equal to $p-1$ and $p$ respectively.
Since $\xi^* < 0$ it is easy to see how only positive terms of the series
survive. 

Collecting all, we eventually find the complete power expansion for 
$R_{1RSB}^{K-SAT}$:
\begin{eqnarray}
& & \frac{\alpha}{\beta m} 
\sum_{l\ge 1}\frac{m^l}{l} 
\sum_{k_1,...,k_l}^{1,\infty} (-\xi^*)^{\sum_{s=1}^l k_s}
\prod_{s=1}^l 
\left(
 \frac{\prod_{r=1}^{k_s-1}(r-m)}{k_s!} 
\right)\times\nonumber\\
& &\Omega^{(l)}
\left[
(1 + {\bf Q}(k_1,...,k_l))^p - p (1 + {\bf Q}(k_1,...,k_l)) {\bf A}(k_1,...,k_l)^{p-1} + (p-1) {\bf A}(k_1,...,k_l)^p
\right] 
\label{ultimoomostro}
\end{eqnarray}
where we have defined
\begin{equation}
{\bf A}(k_1,...,k_l) \equiv 
\left\langle \prod_{s=1}^l 
\left\langle 
\left(
1 + J \tanh (\beta g )
\right)^{k_l}
\right\rangle_g
\right\rangle_G
\end{equation}
Again, every term of the expansion is positive for even $p$ and for
$p$ odd under condition $ 1+{\bf Q}(k_1,...,k_l) \ge 0$.

\section{Appendix C}
Let us briefly sketch the proof of the existence of the thermodynamic
limit of 
free-energy of
the $p$ spin model for $p$.  Let us define a model which
interpolates between two non interacting systems with $N_1$ and $N_2$
spins respectively, and a system of $N=N_1+N_2$ spins. Each clause
$\mu=1,...,M$ will belong to the total system with probability 
$t$, to the first subsystem with probability $N_1/N (1-t)$ and to the 
second subsystem with probability $N_2/N (1-t)$.  
We chose the indices $i_1^\mu,...,i_p^\mu$ in the
following way: for each clause the indices will be i.i.d.  with
probability $t$, the indices will be chosen uniformly in the set
$\{1,...,N\}$, with probability $(1-t)N_1/N$ the indices will be
chosen in $\{1,...,N_1\}$ and with probability $(1-t)N_2/N$ in the
set $\{N_1+1,...,N\}$. Let us consider the free-energy
$F_N(t)=\frac{-1}{N\beta}\log Z(t)$.
A direct calculation of its
$t$-derivative 
\begin{equation}
\frac {dF_N(t)}{dt}=
-\frac 1 \beta \left[
\frac{1}{N^{p}} \sum_{i_1,...,1_p}^{1,N}+
\frac {N_1}{ N} \frac{1}{N_1^p} 
\sum_{i_1,...,1_p}^{1,N_1}+
\frac {N_2} {N} \frac{1}{N_2^{p}}
\sum_{i_1,...,1_p}^{N_1+1,N}
\right]
E 
\langle \log(1+\tanh(\beta
J) \omega(S_{i_1}...S_{i_p}))\rangle_J. 
\end{equation}
Expanding the logarithm in series, observing that thanks to the
symmetry of the $J$ distribution the odd term vanish, introducing the
replica measure and using the convexity of the function $x^p$ for even
$p$ one proves that $\frac {dF_N(t)}{dt}\le 0$ which implies
sub-additivity $F_N\le \frac {N_1}{ N} F_{N_1}+\frac {N_2}{ N}
F_{N_2}$; this is 
in turn is a sufficient condition to the existence of the free-energy
density. The same prove applies to the even $p$ random K-SAT
model. For odd $p$ we face a difficulty similar to the one in the
replica bounds. We can not prove sub-additivity due to the need to
consider negative values of the overlaps, and non convexity of $x^p$
for negative $x$.

\end{document}